\documentclass{PoS}

\usepackage{amsmath}
\usepackage{epsf}

\def\beq{\begin{equation}}
\def\eeq{\end{equation}}
\def\bea{\begin{eqnarray}}
\def\eea{\end{eqnarray}}
\def\beqa{\begin{equation}\begin{array}{l}}
\def\eeqa{\end{array}\end{equation}}
\def\eqlab#1{\label{eq:#1}}
\def\figlab#1{\label{fig:#1}}
\def\tablab#1{\label{tab:#1}}

\def\Eqref#1{Eq.~(\ref{eq:#1})}

\def\fref#1{\ref{fig:#1}}
\def\Figref#1{Fig.~\ref{fig:#1}}
\def\tabref#1{\ref{tab:#1}}

\def\sla#1{#1 \!\!\!\! \slash\,}

\def\slad{\partial \hspace{-2.1mm} \slash}

\def\slaa{a \hspace{-2mm} \slash}

\def\half{\mbox{\small{$\frac{1}{2}$}}}

\def\barr{\left(\begin{array}{c}}
\def\earr{\end{array}\right)}
\def\bmat{\left(\begin{array}{cc}}
\def\emat{\end{array}\right)}
\def\al{\alpha}
\def\be{\beta}
\def\ga{\gamma} 
 \def\De{\Delta}\def\vDe{\varDelta}
\def\veps{\varepsilon}  \def\eps{\epsilon}

\def\la{\lambda}

\def\si{\sigma} 
  
\def\w{\omega}

\def\pa{\partial}

\def\pa{\partial}

\def\cO{\mathcal{O}}

\def\lag{{\mathcal L}}

\def\mathscr{\mathcal}

\def\3d{3-D}

\def\ol#1{\overline{#1}}

\title{Predictions of chiral perturbation theory for Compton scattering off protons}

\ShortTitle{Predictions of chiral perturbation theory for Compton scattering off protons}

\author{\speaker{Vadim LENSKY}%

\\

        ECT* Trento\\

        E-mail: \email{lensky@ect.it}}

\author{Vladimir PASCALUTSA\\

        Institut f{\"u}r Kernphysik, Johannes Gutenberg Universit{\"a}t, Mainz D-55099, Germany \\
}

\abstract{We compute the Compton scattering off the nucleons in the
framework of manifestly covariant baryon chiral perturbation theory (B$\chi$PT).
The results for observables differ substantially from the corresponding calculations
in heavy-baryon chiral perturbation theory (HB$\chi$PT), most appreciably in
the forward kinematics. We verify that the covariant $p^3$ result
fulfills the forward-Compton-scattering sum rules. We also explore the effect of 
the $\Delta$(1232) resonance at order $p^4/\varDelta$, with $\varDelta\approx 300$ MeV,
the resonance excitation energy.
We find that the substantial effect of the $\Delta$-excitation on the nucleon polarizabilities
can naturally be accommodated in the 
manifestly covariant calculation.}

\FullConference{International Workshop on Effective Field Theories: from the pion to the upsilon\\

                 2-6 February 2009\\

                 Valencia, Spain}

\begin{document}

\section{Introduction}

Compton scattering off nucleons
allows to study the structure
and the e.m.\ properties of the nucleon.  
At very low energies, the process depends
on the static e.m.\ moments of nucleon, the 
charge and magnetic moment~\cite{Low:1954kd}.

At larger energies (around 100 MeV and above), the effects of the nucleon structure
 can be detected~\cite{Federspiel:1991yd,Zieger:1992jq,Hal93,MacG95,MAMI01,Drechsel:2002ar,Schumacher:2005an}. 
For instance, these effects show up in the values of
nucleon {\em polarizabilities} --- see the PDG column of Table~\tabref{albe}.

More insights come from chiral
perturbation theory ($\chi$PT), an effective theory
of the low-energy strong interaction~\cite{Weinberg:1978kz,Gasser:1983yg}.
The leading-order $\chi$PT result for the nucleon
polarizabilities is a prediction --- the
low-energy constants (LECs) start to contribute at the next order.
The first calculation of polarizabilities in $\chi$PT~\cite{Bernard:1991rq}
at leading order yields the values shown in the
B$\chi$PT ${\cal O}(p^3)$ column of Table~\tabref{albe}. This calculation was done in
manifestly Lorentz-covariant baryon $\chi$PT (B$\chi$PT), to distinguish it from the heavy-baryon $\chi$PT
 (HB$\chi$PT), which was introduced~\cite{JeM91a} in order to cure
the chiral power counting problems that B$\chi$PT had apparently
had~\cite{GSS89}. Incidentally, the HB$\chi$PT result
agrees with experiment much better, see Table~\tabref{albe}. 
More detailed analyses of Compton scattering in HB$\chi$PT
followed~\cite{BKM,McGovern:2001dd,Beane:2004ra}.
\begin{table}[hbt]
\begin{center}
\begin{tabular}{||c|c|c||c|c||c||}
\hline\hline
&  \multicolumn{2}{|c||}{HB$\chi$PT} & 
\multicolumn{2}{|c||}{B$\chi$PT} & PDG \\
\cline{2-5} 
 &  ${\cal O}(p^3)$ \cite{BKM}&  ${\cal O}(\eps^3) $ \cite{Hemmert:1996rw}& $ {\cal O}(p^3)$ \cite{Bernard:1991rq}
  &$ {\cal O}(p^4/\vDe) $&  \cite{PDG2006}
 \\
\hline
$\alpha^{(p)}$  & 12.2 & 20.8 & 6.8 & 10.8 & $12.0 \pm 0.6$\\
$\beta^{(p)}$ & 1.22 & 14.7 & $-1.8$ & 2.9 & $1.9\pm 0.5 $ \\ 
\hline\hline
\end{tabular}
\end{center}
\caption{The electric ($\al$) and magnetic ($\be$) polarizabilities
of the proton in units of $10^{-4}\,$fm$^3$. The last column quotes
the PDG compilation of experimental results, while
the first two represent the predictions of
the HB$\chi$PT and B$\chi$PT, respectively.
}
\tablab{albe}
\end{table}
However, it was shown more recently~\cite{Gegelia:1999gf,Fuchs:2003qc}
 that B$\chi$PT does not have a problem with power counting {\it per~se}. 
It was also pointed out~\cite{BL99} that the difference
between B$\chi$PT and HB$\chi$PT results can be large due
to the presence of physical cuts and other non-analytic structures.
Moreover, HB$\chi$PT is incompatible with the sum rules~\cite{Lvov:1993ex,Pascalutsa:2004ga,Holstein:2005db}.
Finally, the effect of the $\Delta$ excitation in Compton scattering
cannot be accommodated
in the HB framework in a natural way~\cite{Hildebrandt:2003fm, Pascalutsa:2003zk}
(see also the HB$\chi$PT column of Table~\tabref{albe}).
These observations make a strong case for adopting the B$\chi$PT formalism in favor
of the heavy-baryon one. Here we present the results of a calculation of Compton scattering in B$\chi$PT to orders $p^3$ and $p^4/\De$~\cite{Lensky:2008re}.

\section{Chiral loops and Lagrangians}

Up to $\cO(p^3)$, the $\chi$PT expansion for the Compton amplitude contains the Born graphs,
the Wess-Zumino-Witten anomaly (see, e.g., Ref.~\cite{Pascalutsa:2003zk}), and 
the loop graphs shown in \Figref{loops}.
To calculate these loops, we consider the leading-order chiral Lagrangian for the nucleon:
\beq
\lag^{(1)}_N = \ol N\,( i \sla{D} -{M}_{N} +  g_A \,  
\slaa\,\ga_5 )\, N\,,
\eqlab{Nlagran}
\eeq
where $N$ denotes the isodoublet Dirac field of the nucleon,
$M_N$ is the nucleon mass and $g_A$ is the axial-coupling 
constant, and the chiral covariant derivative is given by
$
D_\mu N = \pa_\mu N  + i v_\mu N \,,
$
whereas the vector and axial-vector fields above are defined 
in terms of the pion field, $\pi^a(x)$, as
\beq
v_\mu  \equiv  \half\, \tau^a v_\mu^a(x) = \frac{1}{2i} \left(u \,
\pa_\mu u^\dagger+u^\dagger \pa_\mu  u \right), \qquad
a_\mu \equiv  \half \,\tau^a a^{\,a}_\mu(x) =
  \frac{1}{2i} \left(u^\dagger \,
\pa_\mu  u- u \,\pa_\mu  u^\dagger \right) , 
\eeq
with $u=\exp(i\pi^a \tau^a/2f )$, and $f$ the pion decay constant.
Then, we apply a redefinition of the nucleon field, $N\to \xi N$, where $\xi$ has the form:
\beq
\xi = \exp\left(\frac{ig_A\,\pi^a \tau^a}{2f}\ga_5 \right)\,.
\eeq
For the one-loop contributions to Compton scattering
it is sufficient to expand up to the second
order in the pion field.
After the expansion, the redefined Lagrangian takes the following form:
\bea
\!\!\!\!\!\!\!\!\!\!\!{\lag'}_N^{(1)} \!& = & \! \ol N\!\left(\! \!i \slad\! -\!{M}_{N}\! 
-i\, \frac{ g_A}{f} M_N \tau^a\pi^a\ga_5 +  \frac{g_A^2}{2f^2} M_N \pi^2
-\frac{(g_A-1)^2}{4f^2} \tau^a\veps^{abc}  \pi^b\slad \pi^c
\right)\! N
 + O(\pi^3)\,.
\eqlab{expNlagran2}
\eea
Finally, one gets the set of diagrams in \Figref{loops}
with the couplings from \Eqref{expNlagran2} instead of the usual set~\cite{Bernard:1991rq}. However, the two sets
of one-loop diagrams give identical expressions for the Compton amplitude. This fact also explains why
the one-loop result for polarizabilities in the linear sigma model with heavy $\si$-meson~\cite{Metz:1996fn}
is exactly the same as in B$\chi$PT at $\cO(p^3)$~\cite{Bernard:1991rq}. 

The $\De$ excitation starts to contribute at order $p^4/\vDe$, where $\vDe=M_\De-M_N\approx 0.3$ GeV. The relevant diagrams are shown in \Figref{loopsD}. The $\Delta$ Born contribution is calculated in the same way as in Ref.~\cite{Pascalutsa:2003zk}, except that we use the values of the $\ga N\to \De$ couplings ($g_M=2.95$ and $g_E=-1.0$) from
the pion-photoproduction analyses of Refs.~\cite{Pascalutsa:2005ts,Pascalutsa:2006up}, and 
also include the corresponding crossed graph. We use $h_A=2.85$, corresponding to the $\De \to \pi N$ decay width of 115~MeV.

The one-particle-reducible graphs in 
Figs.~\fref{loops} and \fref{loopsD} contribute to the nucleon 
mass, field, charge, and magnetic moment renormalization. We adopt the
on-mass-shell renormalization scheme, and use the following values
of the parameters:  $e^2/4\pi = 1/137$, $g_A=1.267$, $f=f_\pi = 0.0924$ GeV,
$m_\pi = 0.139$ GeV, $M_N = 0.9383$ GeV, $\kappa_N = 1.79$ for the proton. 

 \begin{figure}[h]
\centerline{\epsfclipon   \epsfxsize=9.0cm%
  \epsffile{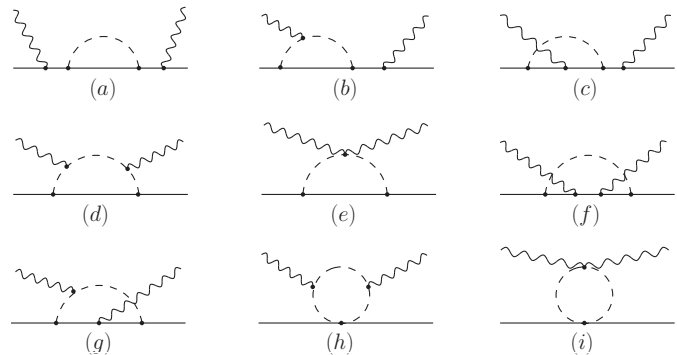} 
}
\caption{
The loop graphs evaluated in this work. Graphs obtained from these by
crossing and time-reversal are not shown, but are evaluated too. 
}
\figlab{loops}
\end{figure}
\begin{figure}[ht]
\centerline{\epsfclipon   \epsfxsize=9.0cm%
  \epsffile{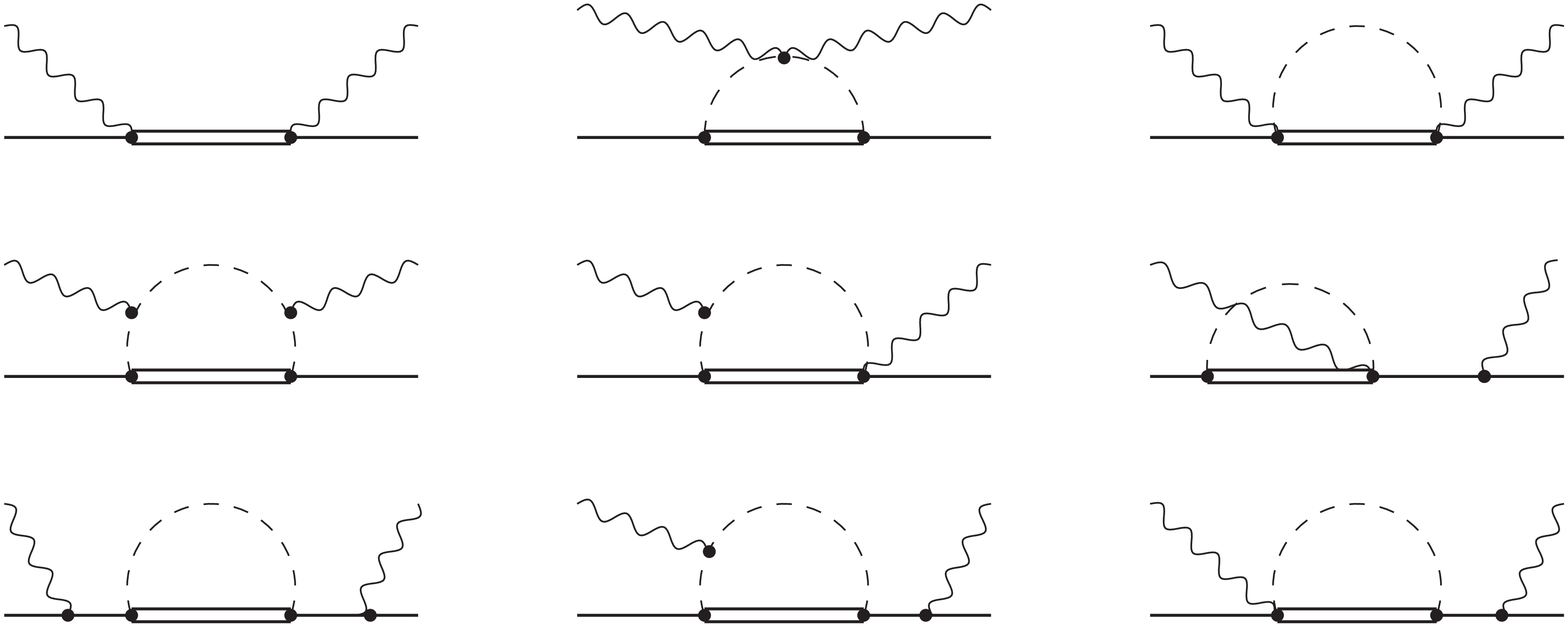} 
}
\caption{
The graphs with $\Delta(1232)$ that contribute at order $p^4/\vDe$. 
Graphs obtained from these by
crossing are not shown, but are calculated as well.
}
\figlab{loopsD}
\end{figure}
\section{Consistency with sum rules}

The amplitude of forward Compton scattering can be related to 
an integral over energy of the photoabsorption cross-section, which in combination
with the low-energy expansion yields a number of model-independent sum rules.
A famous example is the Baldin sum rule:
\beq
\alpha+\beta=\frac{1}{2\pi^2}\int\limits^\infty_0
 d\omega\frac{\sigma_T(\omega)}{\omega^{2}-i0}\eqlab{BSR}
\eeq
that relates the sum of polarizabilities to an integral of the total photoabsorption cross-section $\sigma_T$.

In general, the forward Compton-scattering amplitude can be 
decomposed into two scalar functions of single variable:
\beq
T_{fi}(\omega)=\vec\epsilon^{\,\prime*}\cdot\vec\epsilon \,f(\omega)
+i\vec\sigma\cdot(\vec\epsilon^{\,\prime*}\times\vec\epsilon\,)\,\w\,g(\omega),
\eeq
where $\vec\epsilon^{\,\prime},\ \vec\epsilon$ are the polarization vectors of the initial and final photons, respectively,
 and $\vec\sigma$ are the Pauli spin matrices.
Using analyticity and unitarity,
one can write down the following 
sum rules:
\bea
\!\!\!f(\omega) \!=\! f(0)+\frac{\omega^2}{2\pi^2}\int\limits^\infty_0
 d\omega^\prime\frac{\sigma_T(\omega^\prime)}{\omega^{\prime\,2}-\omega^2-i0}\eqlab{fdis}\,,\quad
g(\omega) \!= \!\frac{1}{4\pi^2}\int\limits^\infty_0
 d\omega^\prime\,\omega^\prime\, 
\frac{\sigma_{1/2}(\omega^\prime)-\sigma_{3/2}(\omega^\prime)}{\omega^{\prime\,2}-\omega^2-i0}\eqlab{gdis}\,,
\eqlab{sumrules}
\eea
where $\sigma_\lambda$ is
the doubly-polarized photoabsorption cross-section, with $\la$ being the helicity
of the initial photon-nucleon state. We showed that 
the loop contributions in \Figref{loops} 
fulfill the sum rules \Eqref{sumrules}. From the sum rules,
one can see that at $\cO(p^3)$, chiral symmetry is not relevant for the forward Compton amplitude.
The graphs $(h)$ and $(i)$ in \Figref{loops} take the role
of chiral symmetry. In the forward angles these graphs vanish,
but play an important role in the backward angles. Without them
the value of $\al -\be$ would diverge as $1/m_\pi^{2}$ in the chiral limit (instead of $1/m_\pi$ as it should). 
Thus, chiral symmetry plays a more prominent role in the backward Compton scattering.

The results for the polarizabilities are in worse agreement
with experiment than the HB$\chi$PT $p^3$ result (see Table~\tabref{albe}).  
This, in fact, opens a room for the
$\De$(1232) contributions. The $\De$(1232) plays 
an important role in nucleon polarizabilities, as can be seen
from the Baldin sum rule and the fact that the 
photoabsorption cross-section is dominated, at lower energies,
by the $\De$ resonance.
In contrast, the HB$\chi$PT $p^3$  value for $\alpha+\beta$ saturates the sum rule, leaving
no room for other contributions.

\section{Results for observables}
In \Figref{fixE}, we show the unpolarized differential cross-section
of the $\gamma p\to\gamma p$ process 
as a function of the scattering angle in center-of-mass system, with
the incident photon energy fixed at just below the pion-production 
threshold. 
The major differences between the HB$\chi$PT and B$\chi$PT $p^3$ calculations
arise at forward angles. This is because at low energies the 
$p^3$ contribution to the cross-section at forward (and backward) angles
is determined by the $p^3$ contribution to $\al+\be$ (and $\al-\be$).
The sum of polarizabilities differs between the two calculations
much more than their difference, and this fact reflects itself in the cross-section.
 
The red solid line with an error band in \Figref{fixE} shows the
result of adding the $\De$ contribution to the covariant $p^3$
result. The $\De$ contribution in B$\chi$PT 
is compatible with both photoproduction
and Compton scattering data, which is further demonstrated 
in \Figref{fixA}, where the $\gamma p\to\gamma p$ cross-section
is plotted as a function of photon energies at fixed angles (in the lab
system).  The HB$\chi$PT result is omitted here, but
can be found in Ref.~\cite{Beane:2004ra}. 
The results for the proton polarizabilities, complete up to ${\cal O}(p^4/\vDe)$,
are displayed in Table~\tabref{albe}.

\begin{figure}[bth]
\centerline{ \epsfclipon  
\epsfxsize=7.0cm%
  \epsffile{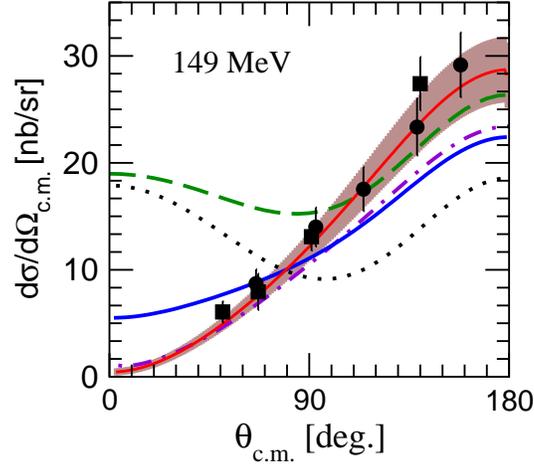} 
}
\caption{(Color online)  Angular dependence of 
the $\ga p\to \ga p$ differential cross-section in
the center-of-mass system for fixed photon-beam energy, 
$E_\gamma^{(lab)}=149$ MeV. Data points are from SAL~\cite{Hal93} ---
filled squares, and MAMI~\cite{MAMI01} --- filled circles. The curves are:
Klein-Nishina --- dotted, Born graphs and WZW-anomaly --- green dashed,
adding the HB$\chi$PT --- violet dash-dotted, adding the B$\chi$PT
--- blue solid. The result of adding the $\Delta$-excitation
contribution to the  B$\chi$PT $p^3$ is shown by the red solid line with a band.}
\figlab{fixE}
\end{figure}
\begin{figure}[bth]
\centerline{\epsfclipon  \epsfxsize=15cm%
  \epsffile{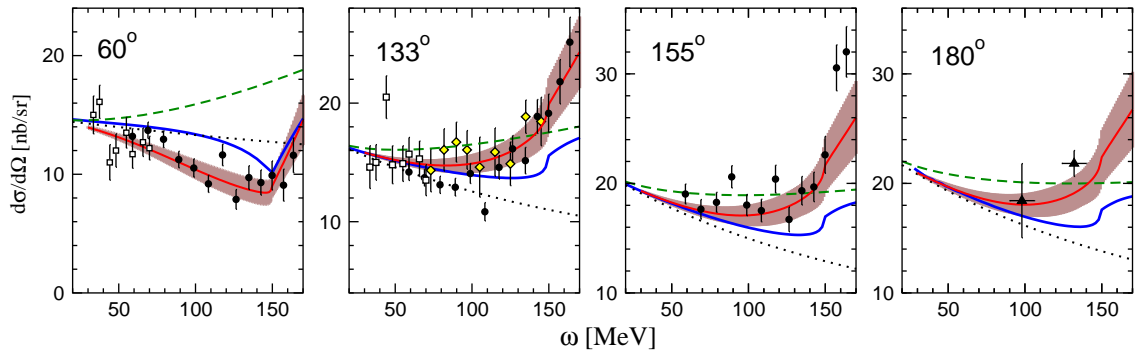} 
}
\caption{(Color online) Energy dependence of 
 the $\ga p\to \ga p$ differential cross-section in
 the laboratory frame for fixed values of the scattering angle. 
 Data: Illinois~\cite{Federspiel:1991yd} --- open squares,
 MAMI~\cite{Zieger:1992jq} --- filled triangles, SAL~\cite{MacG95} ---
 open diamonds, and MAMI~\cite{MAMI01} --- filled circles. The curves are as in Fig.~3.}
\figlab{fixA}
\end{figure}

\section{Conclusion}
We have studied the nucleon Compton scattering in the
framework of B$\chi$PT at orders $p^3$ and $p^4/\vDe$.  
The covariant $p^3$ result fulfills the forward-Compton-scattering sum rules. Chiral symmetry has no effect on
the forward scattering but plays an important 
role at the backward scattering.
For the $\gamma p\to \gamma p$ cross sections we find that the difference between the HB$\chi$PT and B$\chi$PT
results can indeed be unnaturally large, especially in the forward kinematics. We argue that higher-order effects of the  $\Delta$(1232) excitation can more naturally be accommodated in the B$\chi$PT calculation.
This is due to partial cancellation of the relativistic and $\De$-excitation effects which
is explicit in the covariant calculation.
In contrast to the HB$\chi$PT approach, in B$\chi$PT the effect of $\Delta$(1232) appears to be
compatible both with the Compton scattering and pion photoproduction data.

\section*{Acknowledgments}
This work is partially supported  by the European Community Research Infrastructure Activity under the FP6 
"Structuring the European Research Area" programme (HadronPhysics, contract RII3-CT-2004-506078).

\end{document}